\documentclass[twocolumn,showpacs,superscriptaddress,preprintnumbers,nofootinbib,prd]{revtex4}
\usepackage{amsmath,amssymb}
\usepackage{graphicx}
\usepackage{dcolumn}
\usepackage{epsfig,color}
\usepackage{bm}
\usepackage{verbatim}
\def\be{\begin{equation}}
\def\ee{\end{equation}}

\begin{document}

\title{Spectroscopy of the hidden-charm $[qc][\bar q \bar c]$ and $[sc][\bar s \bar c]$ tetraquarks}

\author{Muhammad Naeem Anwar}\email[]{naeem@itp.ac.cn}
\affiliation{CAS Key Laboratory of Theoretical Physics, Institute of Theoretical Physics, Chinese Academy of Sciences, Beijing 100190, China}
\affiliation{University of Chinese Academy of Sciences, Beijing 100049, China}

\author{Jacopo Ferretti}\email[]{jak.ferretti@gmail.com}
\affiliation{CAS Key Laboratory of Theoretical Physics, Institute of Theoretical Physics, Chinese Academy of Sciences, Beijing 100190, China}

\author{Elena Santopinto}\email[]{santopinto@ge.infn.it}
\affiliation{INFN, Sezione di Genova, via Dodecaneso 33, 16146 Genova, Italy}

\begin{abstract}
We calculate the spectrum of $q\bar q c \bar c$ and $s\bar s c \bar c$ tetraquarks, where $q$, $s$ and $c$ stand for light ($u,d$), strange and charm quarks, respectively, in a relativized diquark model, characterized by one-gluon-exchange (OGE) plus confining potential.
In the diquark model, a $q\bar q c \bar c$ ($s\bar s c \bar c$) tetraquark configuration is made up of a heavy-light diquark, $q c$ ($sc$), and anti-diquark, $\bar q \bar c$ ($\bar s \bar c$).
According to our results, 13 charmonium-like observed states can be accommodated in the tetraquark picture, both in the hidden-charm ($q\bar q c \bar c$) and hidden-charm hidden-strange ($s\bar s c \bar c$) sectors.
\end{abstract}


\maketitle

\section{Introduction}
For a few decades after the formulation of the quark model, it was believed that baryons and mesons could be described as the bound states of three valence quarks and a constituent quark-antiquark pair, respectively.
The classification of ground-state hadrons could be easily carried out in terms of group theoretical techniques and the quark model formalism, while resonances might be sorted by making use of effective potentials to describe the spatial excitations related to the inter-quark motion. For example, see Refs.~\cite{Greiner:1989eu,Richard:1992uk,Buchmuller:1992zf,Capstick:2000qj}.

However, more recent data from both $e^+ e^-$ and hadron colliders shed light on hadrons which do not fit well into this standard picture.
They are the so-called {\it exotic XYZ hadrons}, namely multiquark states (tetraquarks and pentaquarks) and particles including gluonic degrees of freedom (hybrids and glueballs).
We are especially interested in tetraquarks, which are mesons containing two valence quarks and two antiquarks.
Among tetraquark candidates, we can mention $Z_c(3900)$ \cite{Ablikim,Liu}, $Z_c(4020)$ \cite{Ablikim:2013emm,Ablikim:2013wzq}, $Z_b(10610)$, $Z_b(10650)$ \cite{Bondar}, and the well-known $X(3872)$ \cite{Choi:2003ue}.
The tetraquark nature is still unclear and several different interpretations have been proposed.
They include: a) Tightly bound objects, just as in the case of normal hadrons, but with more constituents \cite{Jaffe:1976ih,Barbour:1979qi,Weinstein:1983gd,SilvestreBrac:1993ss,Brink:1998as,Maiani:2004vq,Barnea:2006sd,Santopinto:2006my,Ebert:2008wm,Deng:2014gqa,Zhao:2014qva,Anwar:2017toa}; b) Hadro-quarkonia (hadro-charmonia) \cite{Dubynskiy:2008mq,Voloshin:2013dpa,Li:2013ssa,Wang:2013kra,Brambilla:2015rqa,Panteleeva:2018ijz,Ferretti:2018kzy}; c) Loosely bound meson-meson molecules similar to the deuteron \cite{Weinstein:1990gu,Manohar:1992nd,Tornqvist:1993ng,Swanson:2003tb,Hanhart:2007yq,Thomas:2008ja,Baru:2011rs,Valderrama:2012jv,Guo:2013sya,Kang:2016jxw}; d) The result of kinematic or threshold effects caused by virtual particles \cite{Heikkila:1983wd,Pennington:2007xr,Danilkin:2010cc,bottomonium,Lu:2016mbb}; e) The rescattering effects arising by anomalous triangular singularities \cite{Guo:2014iya,Szczepaniak:2015eza,Liu:2015taa}. More details on the previous interpretations can be found in Refs.~\cite{Ali:2017jda,Esposito:2016noz,Olsen:2017bmm,Karliner:2017qhf,Chen:2016qju,Guo:2017jvc,Lebed:2016hpi}. Here, we focus on the first one.

Four quark states can in principle be bound by one-gluon-exchange (OGE) forces. However, their possible emergence and stability is controversial because of the lack of reliable and univocal experimental data. As a consequence, tetraquark model predictions are strongly model dependent and rely on the choice of a specific Hamiltonian, and also on the parameter fitting procedure.
Despite of this, the tetraquark hypothesis is worth to be investigated.

It is worth noting that in the tetraquark hypothesis one obtains a four-quark spectrum which is richer than those generated by molecular models or the inclusion of dynamical/threshold effects in the quark model formalism.
In particular, in molecular models one only expects to get bound states in the proximity of meson-meson decay thresholds; radial excitations cannot take place because of the smallness of meson-meson binding energies.
On the contrary, if one includes threshold effects in the quark model formalism, one gets radial excitations, but exotic charged states of the type $q \bar q Q \bar Q$, where $Q$ is a heavy quark, are forbidden.
A comparison between theoretical predictions for the spectrum and main decay modes of four-quark states and the existing experimental data may allow to distinguish between the previous hypotheses.
The possible emergence of the fully-heavy $QQ \bar Q \bar Q$ bound states may provide a strong indication in favor of the tetraquark one \cite{Anwar:2017toa,Karliner:2016zzc,Berezhnoy:2011xn,Richard:2017vry,Chen:2016jxd,Wang:2017jtz}.

It is also worth to remind that the heavy-light tetraquarks in a $QQ \bar{q}\bar{q}$ configuration are also of considerable interest.\footnote{All the other possible heavy-light tetraquarks can decay strongly by annihilating at least a quark-antiquark pair of the same flavor.}
It would be very interesting to test the possibility of a $QQ \bar{q}\bar{q}$ tetraquark that remains stable against strong decays, but unfortunately there is no experimental evidence yet. Theoretically, $QQ \bar{q}\bar{q}$ was first shown to be stable against strong decays by Lipkin~\cite{Lipkin:1986dw} and Ader \textit{et al.}~\cite{Ader:1981db} long ago.
Very recently, $bb \bar{q} \bar{q}$ was shown to be stable against strong decays but not its charm counterpart $cc \bar{q} \bar{q}$, nor the mixed (beauty+charm) $bc \bar{q} \bar{q}$ state~\cite{Karliner:2017qjm,Eichten:2017ffp}.
For the detailed discussions on the stability of different heavy-light tetraquarks, we refer to the following recent studies~\cite{Czarnecki:2017vco,Luo:2017eub}.

In this paper, we compute the spectrum of $q \bar q c \bar c$ ($q = u,d$) and $s \bar s c \bar c$ tetraquarks.
The calculations are performed within a diquark-antidiquark relativized model, characterized by a one-gluon-exchange potential. The effective degree of freedom of diquark describes two strongly correlated quarks, with no internal spatial excitation.

The tetraquark spectrum is obtained in a two-step process. First of all, the diquark masses are obtained by solving the Schr\"odinger equation with the relativized quark-quark potential \cite{Godfrey:1985xj}. In a second stage, the tetraquark spectrum is calculated by means of the relativized diquark-antidiquark potential \cite{Anwar:2017toa}.
Finally, by comparing our results to the data, we are able to provide some tentative assignments to $XYZ$-type states, including $X(3872)$, $Z_{\rm c}(3900)$, $Z_{\rm c}(4020)$, $Z_{\rm c}(4240)$, $Z_{\rm c}(4430)$, $Y(4008)$, $Y(4260)$, $Y(4360)$, $Y(4630)$ and $Y(4660)$ in the $q\bar q c \bar c$ sector, plus $X(4140)$, $X(4500)$ and $X(4700)$ in the $s\bar s c \bar c$ sector.
The next step of our study of fully- \cite{Anwar:2017toa} and doubly-heavy tetraquarks will be an analysis of the ground-state energies and dominant decay modes, including estimates of the total decay widths and production cross-sections.

The paper is organized as follows. In section~\ref{TQ-Model}, we describe the relativized diquark model and calculation details, and enlist our model parameters.
Section~\ref{resultDiss} is devoted to discussing the results, where we compare our mass predictions both for the $q\bar{q} c\bar{c}$ and $s\bar{s} c\bar{c}$ tetraquarks with the experimental data and those of the previous theoretical studies. We compare our tentative tetraquark assignments for $XYZ$ states with other theoretical interpretations (if available). Finally, we provide a short summary.

\section{Relativized Diquark Model}
\label{TQ-Model}
In a diquark-antidiquark model, the effective degree of freedom of diquark, describing two strongly correlated quarks with no internal spatial excitations, is introduced.
Tetraquark mesons are then interpreted as the bound states of a diquark, $\mathcal D$, and an antidiquark, $\bar {\mathcal D}$.

The ${\mathcal D} - \bar {\mathcal D}$ relative motion is described in terms of a relative coordinate $\bf{r}_{\rm rel}$ (with conjugate momentum ${\bf q}_{\rm rel}$), thus neglecting the internal diquark (antidiquark) structure.
As a result, one turns a four-body problem into a two-body one and gets a spectrum which is less rich than that of a four-body system. Something similar also happens in the baryon sector, where the  spectrum of a quark-diquark system is characterized by a smaller number of states than that of a three quark one. For example, see Refs. \cite{Capstick:1986bm,Capstick:1992th,Bijker:1994yr,Ferraris:1995ui,Ferretti:2011zz,Ferretti:2015ada}.

\subsection{Diquark-antidiquark states}
The diquark (antidiquark) can be found in two different SU$_{\rm c}$(3) color representations, $\bar {\bf 3}_{\rm c}$ (${\bf 3}_{\rm c}$) and ${\bf 6}_{\rm c}$ ($\bar {\bf 6}_{\rm c}$).
As the tetraquark must be a color singlet, there are two possible diquark-antidiquark combinations:
\begin{enumerate}
\item diquark in $\bar {\bf 3}_{\rm c}$, antidiquark in ${\bf 3}_{\rm c}$
\item diquark in ${\bf 6}_{\rm c}$, antidiquark in $\bar {\bf 6}_{\rm c}$
\end{enumerate}
Diquarks (antidiquarks) are made up of two identical fermions and so they have to satisfy the Pauli principle, i.e. the diquark (antidiquark) total wave function,
\begin{equation}
	\Psi_{\mathcal D} = \psi_{\rm c} \otimes \psi_{\rm sf} \otimes \psi_{\rm sp}  \mbox{ },
\end{equation}
where $\psi_{\rm c}$, $\psi_{\rm sf}$ and $\psi_{\rm sp}$ are the color, spin-flavor and spatial wave functions, must be antisymmetric.

Moreover, if for simplicity we neglect the diquarks' internal spatial excitations, their color-spin-flavor wave functions must be antisymmetric.
This limits the possible representations to being only \cite{Jaffe:2004ph,Ferretti:2011zz}
\begin{subequations}
\label{eqn:color-config}
\begin{equation}
  \label{eqn:bar3c}
  \text{color in} ~ \bar {\bf 3}_{\rm c}; ~ \text{symmetric } \psi_{\rm sf}
\end{equation}
and
\begin{equation}
  \label{eqn:6c}
  \text{color in} ~ {\bf 6}_{\rm c}; ~ \text{antisymmetric } \psi_{\rm sf}  \mbox{ }.
\end{equation}
\end{subequations}
In the study of $q\bar q c \bar c$ and $s\bar s c \bar c$ tetraquarks, we consider diquarks (antidiquarks) of the $cq$- and $cs$-type, where $q = u, d$, with isospin $I_{\mathcal D} = \frac{1}{2}$ or 0, respectively. Because of this, for $q\bar q c \bar c$ states both $I = 0$ and $I = 1$ tetraquark isospin combinations are possible, while in the $s\bar s c \bar c$ case one necessarily has $I = 0$.
We can determine the $J^{PC}$ quantum numbers of the tetraquarks by applying the restrictions for the diquark-antidiquark limit, i.e. $L_{\mathcal D}=L_{\bar {\mathcal D}}=0$ and color $\bar {\bf 3}_{\rm c}\otimes \bf 3_{\rm c}$.
This is because we expect that color-sextet diquarks, Eq. (\ref{eqn:6c}), will be higher in energy than color-triplet ones or even that they will not be bound at all \cite{Jaffe:2004ph,Lichtenberg:1996fi,Santopinto:2006my}. Thus, we are left with the (\ref{eqn:bar3c}) diquark representation.
The latter can be further decomposed in terms of the diquark spin and flavor content. As a result, we get a spin-0, flavor-antisymmetric representation, the scalar diquark, and a spin-1, flavor-symmetric representation, the axial-vector diquark.
With these restrictions, the parity of a tetraquark in the diquark-antidiquark limit is
\begin{equation}
	P=(-1)^{L_{\mathcal D \bar{\mathcal D}}}  \mbox{ },
\end{equation}	
while the charge conjugation (obviously only for its eigenstates) is
\begin{equation}
	C=(-1)^{L_{\mathcal D \bar{\mathcal D}}+S_{\rm tot}}  \mbox{ }.
\end{equation}
Here, $L_{\mathcal D \bar{\mathcal D}}$ is the diquark-antidiquark relative orbital angular momentum, and ${\bf S}_{\rm tot} = {\bf S}_{\mathcal D} + {\bf S}_{\bar{\mathcal D}}$.

\subsection{Relativized model Hamiltonian}
\label{Model Hamiltonian}
We consider the following Hamiltonian
\begin{equation}
   \label{eqn:Hmodel}	
   \begin{array}{rcl}
   	\mathcal{H}^{\rm REL} & = & T + V(r_{\rm rel}) \\
   	& = & \sqrt{q_{\rm rel}^2+m_{\mathcal D}^2} + \sqrt{q_{\rm rel}^2+m_{\bar{\mathcal D}}^2} + V(r_{\rm rel})  \mbox{ },
   \end{array}
\end{equation}
where $\sqrt{q_{\rm rel}^2+m_{\mathcal D, \bar{\mathcal D}}^2}$ are the diquark (antidiquark) kinetic energies, with diquark (antidiquark) masses $m_{\mathcal D}$ ($m_{\bar{\mathcal D}}$), and $V(r_{\rm rel})$ the OGE plus confining potential.
The usual form for $V(r_{\rm rel})$ is
\begin{equation}
	\label{eqn:Vr12-old}
	\begin{array}{rcl}
		V(r_{\rm rel}) & = & \left[ \frac{\alpha_{\rm s}}{r_{\rm rel}} - \frac{3 \beta}{4} \mbox{ } r_{\rm rel} - \frac{8 \pi \alpha_{\rm s} \delta({\bf  r}_{\rm rel})}{3 m_{{\mathcal D}} m_{\bar{\mathcal D}}} \mbox{ }
		{\bf S}_{{\mathcal D}} \cdot {\bf S}_{\bar{\mathcal D}} \mbox{ } \right.  \\
		& - &  \left. \frac{\alpha_{\rm s}}{m_{{\mathcal D}} m_{\bar{\mathcal D}} r_{\rm rel}^3} \left(\frac{3{\bf  S}_{{\mathcal D}} \cdot {\bf  r}_{\rm rel}
		{\bf  S}_{\bar{\mathcal D}} \cdot {\bf  r}_{\rm rel}}{r_{\rm rel}^2} - {\bf  S}_{{\mathcal D}} \cdot {\bf  S}_{\bar{\mathcal D}}\right) \right. \\
		& - & \left. \frac{3}{4} \Delta E \right] \frac{\lambda_{{\mathcal D}}^a}{2} \frac{\lambda_{\bar{\mathcal D}}^a}{2} \mbox{ },
	\end{array}
\end{equation}
where $\lambda_{{\mathcal D},{\bar{\mathcal D}}}^a$ are Gell-Mann color matrices, $\Delta E$ a constant, $\alpha_{\rm s}$ the strength of the color-Coulomb interaction, and $\beta$ that of the linear confining potential.

The hyperfine interaction of Eq. (\ref{eqn:Vr12-old}) is an illegal operator in the Schr\"odinger equation; moreover, the Coulomb-like potential should be regularized in the origin \cite{Weinstein:1983gd}.
To overcome these difficulties, we follow the prescriptions of Refs. \cite{Godfrey:1985xj,Celmaster:1977vh,Capstick:1986bm} and re-write Eq. (\ref{eqn:Vr12-old}) as
\begin{subequations}
\begin{equation}
	\label{eqn:Vr12-new}
	\begin{array}{rcl}
		V(r_{\rm rel}) & = & \beta r_{\rm rel} + G(r_{\rm rel}) + \frac{2 {\bf S}_{{\mathcal D}} \cdot {\bf S}_{\bar{\mathcal D}}}{3 m_{{\mathcal D}} m_{\bar{\mathcal D}}}
		\mbox{ } \nabla^2 G(r_{\rm rel}) \\
		& - & \frac{1}{3 m_{{\mathcal D}} m_{\bar{\mathcal D}}} \left(3 {\bf S}_{{\mathcal D}} \cdot \hat r_{\rm rel} \mbox{ } {\bf S}_{\bar{\mathcal D}} \cdot \hat r_{\rm rel}
		- {\bf S}_{{\mathcal D}} \cdot {\bf S}_{\bar{\mathcal D}}\right) \\
		& \times & \left(\frac{\partial^2}{\partial r_{\rm rel}^2}
		- \frac{1}{r_{\rm rel}} \frac{\partial}{\partial r_{\rm rel}}\right) G(r_{\rm rel}) + \Delta E \mbox{ },
	\end{array}
\end{equation}
where the Coulomb-like potential is given by \cite{Godfrey:1985xj,Capstick:1986bm}
\begin{equation}
	\label{eqn:G(r)}
	G(r_{\rm rel}) = - \frac{4 \alpha_{\rm s}(r_{\rm rel})}{3 r_{\rm rel}} =
	- \sum_k \frac{4 \alpha_k}{3 r_{\rm rel}} \mbox{ Erf}(\tau_{{{\mathcal D}}{\bar{\mathcal D}}k} r_{\rm rel})  \mbox{ }.
\end{equation}
Here, Erf is the error function \cite{Gradshteyn-Ryzhik} and $\tau_{{{\mathcal D}}{\bar{\mathcal D}}k}$ \cite{Godfrey:1985xj,Capstick:1986bm}
\begin{equation}
	\tau_{{{\mathcal D}}{\bar{\mathcal D}}k} = \frac{\gamma_k \sigma_{{{\mathcal D}}{\bar{\mathcal D}}}}{\sqrt{\sigma_{{{\mathcal D}}{\bar{\mathcal D}}}^2+\gamma_k^2}} \mbox{ },
\end{equation}
with
\begin{equation}
        \begin{array}{l}
	\sigma_{{{\mathcal D}}{\bar{\mathcal D}}} = \sqrt{\frac{1}{2} \sigma_0^2 \left[1 + \left(\frac{4 m_{{\mathcal D}} m_{\bar{\mathcal D}}}{(m_{{\mathcal D}} + m_{\bar{\mathcal D}})^2}\right)^4\right]
	+ s^2 \left(\frac{2 m_{{\mathcal D}} m_{\bar{\mathcal D}}}{m_{{\mathcal D}} + m_{\bar{\mathcal D}}}\right)^2}
	\end{array} \mbox{ }.
\end{equation}
\end{subequations}

The values of the parameters $\alpha_k$ and $\gamma_k$ ($k = 1,2,3$), $\sigma_0$ and $s$, extracted from Refs. \cite{Godfrey:1985xj,Capstick:1986bm}, are given in Table \ref{tab:Model-parameters}.
The value of the $qc$ scalar diquark mass, $M_{qc}^{\rm s}$, is extracted from Ref. \cite{Maiani:2004vq}.
The values of the $qs$ scalar and axial-vector diquark masses, $M_{sc}^{\rm s}$ and $M_{sc}^{\rm av}$, are estimated by binding a $sc$ ($\bar s \bar c$) pair via the OGE plus confining potential~\cite{Godfrey:1985xj,Anwar:2017toa}.
The only free parameters of our calculation are thus the strength of the linear confining interaction, $\beta$, the $qc$ axial-vector diquark mass, $M_{qc}^{\rm av}$, and $\Delta E$ (see Table \ref{tab:Model-parameters}); they are fitted to the reproduction of the experimental data \cite{Nakamura:2010zzi}, as discussed in Sec. \ref{D-aD-spectrum}.

\begin{table}[!htbp]
\centering
\begin{tabular}{ccccccc}
\hline
\hline
Parameter  & \hspace{0.15cm} & Value   & \hspace{1cm} & Parameter  & \hspace{0.15cm} & Value \\
\hline
$\alpha_1$ & & 0.25 $\dag$                  & & $\gamma_1$ & & 2.53 fm$^{-1}$ $\dag$ \\
$\alpha_2$ & & 0.15 $\dag$                  & & $\gamma_2$ & & 8.01 fm$^{-1}$ $\dag$ \\
$\alpha_3$ & & 0.20 $\dag$                  & & $\gamma_3$ & & 80.1 fm$^{-1}$ $\dag$ \\
$\sigma_0$ & & 9.29 fm$^{-1}$ $\dag$ & & $s$ & & 1.55 $\dag$ \\
$\beta$        & & 3.90 fm$^{-2}$   & & $\Delta E$   & & $-370$ MeV \\
$M_{cq}^{\rm s}$ & & 1933 MeV $\dag$ & & $M_{cq}^{\rm av}$ & & 2250 MeV \\
$M_{cs}^{\rm s}$ & & 2229 MeV & & $M_{cs}^{\rm av}$ & & 2264 MeV \\
\hline
\hline
\end{tabular}
\caption{Parameters of the model Hamiltonian of Eq. (\ref{eqn:Hmodel}). The values denoted by the symbol $\dag$ are extracted from previous studies. In the upper part of the table, we give the values of the Coulomb-like potential parameters, $\alpha_1$, $\alpha_2$, $\alpha_3$, $\gamma_1$, $\gamma_2$, $\gamma_3$, $\sigma_0$ and $s$, extracted from Refs. \cite{Godfrey:1985xj,Capstick:1986bm}. The value of $M_{cq}^{\rm s}$ ($q = u,d$) is extracted from Ref. \cite{Maiani:2004vq}; those of $\beta$, $M_{cq}^{\rm av}$ and $\Delta E$ are fitted to the reproduction of the experimental data \cite{Nakamura:2010zzi}. The values of the $qs$ scalar and axial-vector diquark masses, $M_{sc}^{\rm s}$ and $M_{sc}^{\rm av}$, are estimated by binding a $sc$ ($\bar s \bar c$) pair via a OGE plus confining potential~\cite{Godfrey:1985xj,Anwar:2017toa}.}
\label{tab:Model-parameters}
\end{table}

\begin{table*}
  \renewcommand\arraystretch{1.5}
  \centering
  \begin{tabular}{p{2cm}p{2cm}p{2.8cm}p{2cm}p{3.8cm}p{1.8cm}}
  \hline\hline
  State            & $J^{PC}$          & $M_{\textrm{exp}}$ (MeV) & $\Gamma$ (MeV) & Observing Process & Experiment \\ \hline
  $X(3872)$        & $1^{++}$      & $3871.69\pm0.17$         & $<1.7$         & $B^\pm\to K^\pm \pi^+ \pi^- J/\psi$  & Belle \\
  $Z_{c}(3900)$    & $1^{+-}$     & $3886.6 \pm2.4 $         & $28.1\pm2.6$   & $e^+ e^- \to \pi^+ \pi^- J/\psi$   & BESIII \\
  $Y(4008)$        & $1^{--}$        & $4008\pm40$              & $226\pm44$     & $e^+ e^- \to \gamma_{\textrm{ISR}}\pi^+ \pi^- J/\psi$ & Belle\\
  $Z_c(4020)^\pm$  & $1^{+-}$   & $4024.1\pm1.9$             & $13\pm5$      & $e^+ e^- \to \pi^+ \pi^- h_c$ & BESIII \\
  $X(4140)$        & $1^{++}$      & $4146.8\pm2.5$           & $19^{+8}_{-7}$ & $ \gamma \gamma \to \phi J/\psi$   & CDF \\
  $Z_c(4240)^\pm$  & $0^{-}$    & $4239\pm18^{+45}_{-10}$ & $220\pm47^{+108}_{-74}$ & $B^0 \to K^+ \pi^- \psi(2S)$ & LHCb \\
  $Y(4260)$        & $1^{--}$        & $4230\pm8$               & $55\pm19$      & $e^+ e^- \to \gamma_{\textrm{ISR}}\pi^+ \pi^- J/\psi$& BaBar \\
  $X(4274)$        & $1^{++}$      & $4273^{+19}_{-9}$     & $56^{+14}_{-16}$ & $B^+ \to J/\psi \phi K^+$  & CDF, LHCb \\
  $Y(4360)$        & $1^{--}$        & $4341\pm8$               & $102\pm9$      & $e^+ e^- \to \gamma_{\textrm{ISR}}\pi^+ \pi^- \psi(2S)$  & Belle \\
  $Z_c(4430)^\pm$  & $1^{+}$   & $4478^{+15}_{-18}$       & $181\pm31$     & $B \to K \pi^\pm \psi(2S)$   & Belle \\
  $X(4500)$        & $0^{++}$      & $4506^{+16}_{-19}$       & $92\pm29$      & $B^+ \to J/\psi \phi K^+$   & LHCb \\
  $Y(4630)$        & $1^{--}$        & $4634^{+8}_{-7}$         & $92^{+40}_{-24}$& $e^+ e^- \to \Lambda^{+}_{c} \Lambda^{-}_{c}$ & Belle \\
  $Y(4660)$        & $1^{--}$        & $4643\pm9$               & $72\pm11$      & $e^+ e^- \to \gamma_{\textrm{ISR}}\pi^+ \pi^- \psi(2S)$   & Belle \\
  $X(4700)$        & $0^{++}$      & $4704^{+17}_{-26}$       & $120\pm50$     & $B^+ \to J/\psi \phi K^+$   & LHCb \\
  \hline\hline
  \end{tabular}
  \caption{Experimental details on hidden-charm exotica which are discussed in this study.
  The last two columns show the first observation mode and the experiment where the discovery took place, respectively. The enlisted values are taken from the PDG~\cite{Nakamura:2010zzi}.}
  \label{exotica}
\end{table*}

\section{Results and Discussions}
\label{resultDiss}

\subsection{$q\bar q c \bar c$ tetraquark spectrum}
\label{D-aD-spectrum}
\begin{figure}
\begin{center}
\includegraphics[width=7.5cm]{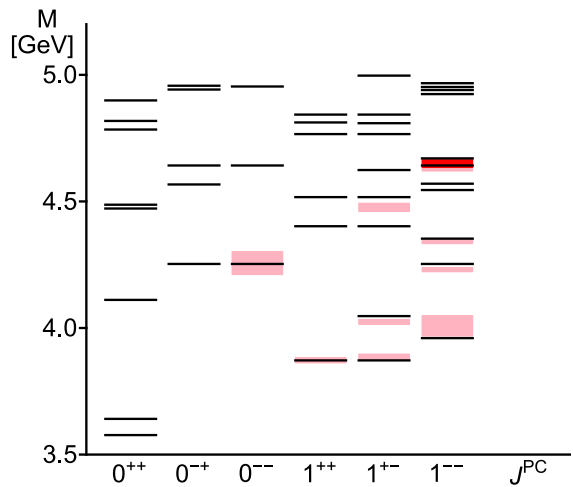}
\end{center}
\caption{The $qc \bar q \bar c$ tetraquark spectrum (lines), obtained by solving the eigenvalue problem of Eq. (\ref{eqn:Hmodel}), is compared to the existing experimental data for $XYZ$ exotics (boxes). For the numerical values, see Tables \ref{exotica} and \ref{tab:tetraquark-spectrum}.}
\label{fig:tetra-spectrum}
\end{figure}

\begin{figure}
\begin{center}
\includegraphics[width=7.5cm]{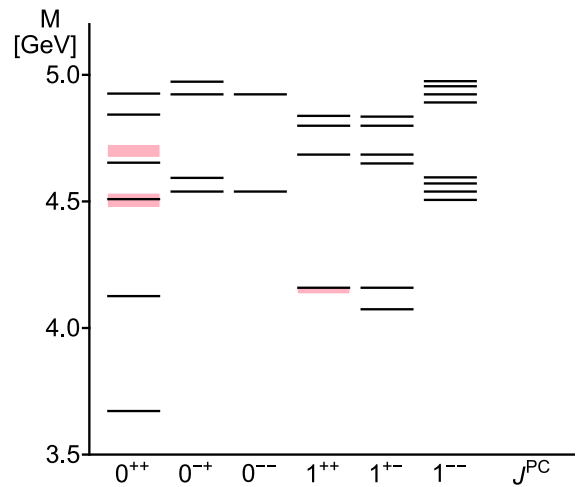}
\end{center}
\caption{As Fig.~\ref{fig:tetra-spectrum}, but for $s\bar s  c \bar c$ tetraquark states.}
\label{fig:tetra-spectrum2}
\end{figure}

In Table \ref{tab:tetraquark-spectrum} and Figs. \ref{fig:tetra-spectrum} and \ref{fig:tetra-spectrum2}, our theoretical predictions for the masses of $q \bar q c \bar c$ ($q = u,d$) and $s \bar s c \bar c$ $0^{++}$, $1^{++}$, $1^{+-}$, $1^{--}$, $0^{--}$ and $0^{-+}$ tetraquark states are compared to the existing experimental data \cite{Nakamura:2010zzi}.
Our results are obtained by solving the eigenvalue problem of the model Hamiltonian [Eq.~(\ref{eqn:Hmodel})] via a numerical variational procedure with harmonic oscillator trial wave functions.
The model parameters, reported in Table \ref{tab:Model-parameters}, are partly extracted from those of previous studies and partly fitted to the reproduction of the spectrum of suspected charmonium-like exotic states \cite{Nakamura:2010zzi,Esposito:2016noz}.

\subsubsection{$X_{\rm c}$ and $Z_{\rm c}$ states}
It is worth noting that we are able to make some clear assignments, as in the case of $X(3872)$, $Z_{\rm c}(3900)$, $Z_{\rm c}(4020)$ and $Z_{\rm c}(4240)$.
This is because the mass difference between the predicted and experimental masses is within the typical error of a quark model calculation, of the order of $30-50$ MeV.

The $X(3872)$, discovered by Belle in $B^\pm\to K^\pm \pi^+ \pi^- J/\psi$ decays \cite{Choi:2003ue}, was the first example of quarkonium-like candidate for a non-standard or exotic meson.
This is a well established meson \cite{Nakamura:2010zzi,Aubert:2004ns,Acosta:2003zx,Abazov:2004kp}, with extremely peculiar features: its mass is $80-100$ MeV below quark model predictions \cite{Godfrey:1985xj} and very close to the $D^0 \bar D^{*0}$ threshold, it is quite narrow ($\Gamma < 1.2$ MeV) and exhibits strong isospin violation in its decays.

In the present study, the $X(3872)$ is interpreted as an $S$-wave scalar diquark, axial-vector antidiquark bound state with $J^{PC} = 1^{++}$ quantum numbers.
This is the same interpretation as Ref. \cite{Maiani:2004vq}, where the authors calculated the spectrum of $c\bar q c \bar q$ tetraquarks by means of an algebraic mass formula, giving the $X(3872)$ mass as input, and Ref. \cite{Ebert:2008wm}, where the authors calculated the tetraquark spectrum in a relativistic diquark-antidiquark model with one-gluon exchange and long-range vector and scalar linear confining potentials.
In the molecular model, the $X(3872)$ is described as a $D^0 \bar D^{*0}$ meson-meson bound state \cite{Tornqvist:1993ng,Swanson:2003tb,Hanhart:2007yq,Thomas:2008ja,Baru:2011rs,Valderrama:2012jv,Guo:2013sya,Kang:2016jxw}.

The $Z_{\rm c}(3900)$ is a charged charmonium-like meson, with $1^{+-}$ quantum numbers, observed at about the same time by BESIII \cite{Ablikim} and Belle \cite{Liu}.
Its exotic quantum numbers and the value of its mass, about 12 MeV above the $D^0 D^{*+}$ threshold, is incompatible with both the charmonium and molecular model interpretations.
In Ref. \cite{Voloshin:2013dpa}, the $Z_{\rm c}(3900)$ was interpreted as a hadro-charmonium state, namely as a $J/\psi$ embedded in an $S$-wave spin-less excitation of the light-quark matter with the quantum numbers of a pion, $J^P = 0^-$.
Our interpretation is the same as Refs. \cite{Maiani:2004vq,Ebert:2008wm}, namely as the $C$-odd partner of the $X(3872)$ \cite[Eq. (22)]{Maiani:2004vq}.

The $Z_{\rm c}(4020)$ was seen by BESIII in a study of $h_{\rm c}(1P) \pi^+ \pi^-$ final states \cite{Ablikim:2013wzq}; its quantum numbers are $J^{PC} = 1^{+-}$.
We interpret the $q \bar q c \bar c$ state of Table \ref{tab:tetraquark-spectrum}, with $1[(1,1)1,0]1$ and $1^{+-}$ quantum numbers, as $Z_{\rm c}(4020)$.
Other interpretations include a $D^* \bar D^*$ molecular state with $1^{+-}$ quantum numbers \cite{Tornqvist:1993ng,Guo:2013sya}, binded by one-pion-exchange and/or contact interactions, or a tightly bound tetraquark configuration \cite{Maiani:2004vq}.

In 2014, LHCb confirmed the existence of the $Z_{\rm c}(4430)$ in $\pi^\pm \psi(2S)$ and, within the same dataset, also observed a lighter and wider structure named the $Z_{\rm c}(4240)$ \cite{Aaij:2014jqa,Aaij:2015zxa}. Further experimental confirmation of the $Z_{\rm c}(4240)$ would be helpful. In our study, we interpret the $Z_{\rm c}(4240)$ as a $P$-wave scalar diquark, axial-vector antidiquark bound state with $J^{PC} = 0^{--}$.

Finally, the $Z_{\rm c}(4430)$ was the first established candidate for a charged charmonium-like meson. It was observed by Belle as a peak in the the invariant mass of the $\psi(2S) \pi^+$ system in $\bar B \rightarrow \psi(2S) \pi^+ K$ \cite{Choi:2007wga}.
In our study, we interpret the $Z_{\rm c}(4430)$ as a $D$-wave scalar diquark, axial-vector antidiquark bound state with $J^{PC} = 1^{+-}$ quantum numbers.
However, in this case the assignment is more dubious, because the experimental mass of the meson falls in the energy interval between the $2[(1,0)1,0]1$ and $1[(1,0)1,2]1$, $J^{PC} = 1^{+-}$ states of Table \ref{tab:tetraquark-spectrum}.
It is worth noting that the $Z_{\rm c}(4430)$ was interpreted as a $2S$ scalar diquark, axial-vector antidiquark bound state in Ref. \cite{Ebert:2008wm}.
Moreover, the $Z_{\rm c}(3900)$ and $Z_{\rm c}(4430)$ were assigned as the ground state and first radial excitation of the same tetraquark with $J^P=1^+$, and several strong decays were explored~\cite{Agaev:2017tzv}.

\begin{table*}
\begin{tabular}{|ccccc||ccccc|}
\hline
\hline
State                       & $J^{PC}$   & $N[(S_{\mathcal D},S_{\bar {\mathcal D}})S,L]J$ & $E^{\rm th}$ & $E^{\rm exp}$ & State & $J^{PC}$   & $N[(S_{\mathcal D},S_{\bar {\mathcal D}})S,L]J$ & $E^{\rm th}$ & $E^{\rm exp}$ \\
($q\bar q c \bar c$) &                   &                                               &           [MeV] &                         [MeV]  & ($s\bar s c \bar c$) &                   &                                            &           [MeV] &                         [MeV] \\
\hline
                               & $0^{++}$    &  $1[(0,0)0,0]0$                      & 3577                           &                                 & & $0^{++}$    &  $1[(1,1)0,0]0$                      & 3672                     &                                  \\
                               & $0^{++}$    &  $1[(1,1)0,0]0$                      & 3641                           &                                  && $0^{++}$    &  $1[(0,0)0,0]0$                      & 4126                     &                                  \\
                               & $0^{++}$    &  $2[(0,0)0,0]0$                      & 4111                            &          &$X(4500)$                 & $0^{++}$    &  $2[(1,1)0,0]0$                      & 4509 & $4506\pm11^{+12}_{-15}$         \\
                               & $0^{++}$    &  $3[(0,0)0,0]0$                      & 4480                          &     & $X(4700)$ & $0^{++}$    &  $2[(0,0)0,0]0$            & 4653           & $4704^{+17}_{-26}$                        \\
                               & $0^{++}$    &  $2[(1,1)0,0]0$                      & 4482                           &                                  && $0^{++}$    &  $3[(1,1)0,0]0$                      & 4926                     &                                   \\
                               & $0^{++}$    &  $4[(0,0)0,0]0$                      & 4784                           &                                  & & $0^{++}$    &  $1[(1,1)2,2]0$                     & 4843                  &                \\
                               & $0^{++}$    &  $1[(1,1)2,2]0$                      & 4818                           &                                  & &      &                        &                            &                \\
                               & $0^{++}$    &  $3[(1,1)0,0]0$                      & 4899                           &                                 & &      &                        &                            &                \\
\hline
$X(3872)$      & $1^{++}$    &  $1[(1,0)1,0]1$                      & 3872                           &  $3871.69\pm0.17$  & $X(4140)$ & $1^{++}$    &  $1[(1,0)1,0]1$          & 4159       &      $4146.8\pm2.5$                       \\
                               & $1^{++}$    &  $2[(1,0)1,0]1$                      & 4402                           &                                  & & $1^{++}$    &  $2[(1,0)1,0]1$                      & 4685                     &                                  \\
                               & $1^{++}$    &  $1[(1,0)1,2]1$                      & 4517                           &                                  & & $1^{++}$    &  $1[(1,0)1,2]1$                      & 4799                     &                                  \\
                               & $1^{++}$    &  $3[(1,0)1,0]1$                      & 4766                           &                                  & & $1^{++}$    &  $1[(1,1)2,2]1$                      & 4838                     &                \\
                               & $1^{++}$    &  $1[(1,1)2,2]1$                      & 4812                            &                                  & &      &                        &                            &                \\
                               & $1^{++}$    &  $2[(1,0)1,2]1$                      & 4843                           &                                  & &      &                        &                            &                \\
\hline
$Z_{\rm c}(3900)$ & $1^{+-}$     &  $1[(1,0)1,0]1$               & 3872                           &  $3886.6\pm2.4$    & & $1^{+-}$     &  $1[(1,1)1,0]1$                      & 4074                     &                                  \\
$Z_{\rm c}(4020)$ & $1^{+-}$     &  $1[(1,1)1,0]1$               & 4047                           &  $4024.1\pm1.9$  & & $1^{+-}$     &  $1[(1,0)1,0]1$                      & 4159                     &                                  \\
                               & $1^{+-}$     &   $2[(1,0)1,0]1$             & 4402                           & &       & $1^{+-}$&   $2[(1,1)1,0]1$                     & 4650                     &                                  \\
$Z_{\rm c}(4430)$   & $1^{+-}$     &   $1[(1,0)1,2]1$            & 4517 & $4478^{+15}_{-18}$ & & $1^{+-}$  &   $2[(1,0)1,0]1$                     & 4685                     &                                  \\
                               & $1^{+-}$     &   $2[(1,1)1,0]1$                     & 4624                           &                                  & & $1^{+-}$     &   $1[(1,0)1,2]1$                     & 4799                     &                                  \\
                               & $1^{+-}$     &    $3[(1,0)1,0]1$                    & 4766                           &                                  & & $1^{+-}$     &  $1[(1,1)1,2]1$                      & 4835                     &                                  \\
                               & $1^{+-}$     &  $1[(1,1)1,2]1$                      & 4809                           &                                  & &      &                        &                            &                \\
                               & $1^{+-}$     &   $2[(1,0)1,2]1$                     & 4843                           &                                  & &      &                        &                            &                \\
                               & $1^{+-}$     &   $3[(1,1)1,0]1$                     & 4997                           &                                 & &      &                        &                            &                \\
\hline
$Y(4008)$   & $1^{--}$      &  $1[(0,0)0,1]1$                      & 3960                           &  $4008\pm40$    & & $1^{--}$      &  $1[(0,0)0,1]1$                      & 4506                     &                                  \\
$Y(4260)$     & $1^{--}$      &   $1[(1,0)1,1]1$                     & 4253                           &  $4230\pm8$            & & $1^{--}$      &   $1[(1,0)1,1]1$                     & 4539                     &                                  \\
$Y(4360)$     & $1^{--}$      &   $2[(0,0)0,1]1$                     & 4353                           &  $4341\pm8$    & & $1^{--}$      &   $1[(1,1)0,1]1$                     & 4571                     &                                  \\
                               & $1^{--}$      &   $1[(1,1)0,1]1$                     & 4545                           &                                  & & $1^{--}$      &   $2[(0,0)0,1]1$                     & 4891                     &                                  \\
                               & $1^{--}$      &   $1[(1,1)2,1]1$                     & 4570                           &                                  & & $1^{--}$      &   $1[(1,1)2,1]1$                     & 4595                           &            \\
$Y(4630)$         & $1^{--}$      &   $2[(1,0)1,1]1$                     &  4642                          &  $4634^{+8}_{-7}$     & & $1^{--}$      &   $2[(1,0)1,1]1$                     & 4923                     &                                  \\
$Y(4660)$     & $1^{--}$      &   $3[(0,0)0,1]1$                     &  4670                          &  $4643\pm9$     & & $1^{--}$      &   $2[(1,1)0,1]1$                     & 4955                     &                                  \\
                               & $1^{--}$      &   $2[(1,1)0,1]1$                     & 4929                           &                                  & & $1^{--}$      &  $2[(1,1)2,1]1$                      & 4975                           &           \\
                               & $1^{--}$      &   $4[(0,0)0,1]1$                     &  4946                          &      & &      &                        &                            &                \\
                               & $1^{--}$      &  $2[(1,1)2,1]1$                      & 4949                          &                                 & &      &                        &                            &                \\
                               & $1^{--}$      &   $3[(1,0)1,1]1$                     &  4954                          &        & &      &                        &                            &                \\
\hline
$Z_{\rm c}(4240)$ & $0^{--}$    &  $1[(1,0)1,1]0$                        & 4253                           &   $4239\pm18^{+45}_{-10}$   & & $0^{--}$    &  $1[(1,0)1,1]0$                        & 4539                     &                        \\
                               & $0^{--}$    &  $2[(1,0)1,1]0$                        & 4642                           &                              & & $0^{--}$    &  $2[(1,0)1,1]0$                        & 4923                     &                                  \\
                               & $0^{--}$    &  $3[(1,0)1,1]0$                        & 4954                           &                                  & &      &                        &                            &                \\
\hline
                               & $0^{-+}$    &  $1[(1,0)1,1]0$                        & 4253                           &                                  & & $0^{-+}$    &  $1[(1,0)1,1]0$                        & 4539                     &                                  \\
                               & $0^{-+}$    &  $1[(1,1)1,1]0$                        & 4567                           &                                  & & $0^{-+}$    &  $1[(1,1)1,1]0$                        & 4593                     &          \\
                               & $0^{-+}$    &  $2[(1,0)1,1]0$                        & 4642                           &                                  & & $0^{-+}$    &  $2[(1,0)1,1]0$                        & 4923                     &                                  \\
                               & $0^{-+}$    &  $2[(1,1)1,1]0$                        & 4947                           &                                  & & $0^{-+}$    &  $2[(1,1)1,1]0$                    & 4973                     &                \\
                               & $0^{-+}$    &  $3[(1,0)1,1]0$                        & 4954                           &                                  & &      &                        &                            &                \\
\hline
\hline
\end{tabular}
\caption{The $q \bar q c \bar c$ ($q = u,d$) and $s \bar s c \bar c$ tetraquark spectrum (up to 5 GeV), obtained by solving the eigenvalue problem of Eq. (\ref{eqn:Hmodel}) with the model parameters of Table \ref{tab:Model-parameters}, is compared to the existing experimental data \cite{Nakamura:2010zzi}. In the third column, we give the quantum numbers of the predicted tetraquark states: $N$ stands for the radial quantum number, $S_{\rm D}$ and $S_{\bar {\rm D}}$ are the spin of the diquark and antidiquark, respectively, coupled to the total spin of the meson, $S$; the latter is coupled to the orbital angular momentum, $L$, to get the total angular momentum of the tetraquark, $J$. For more details on the tetraquark basis, see App.~\ref{Classification of tetraquark states}.}
\label{tab:tetraquark-spectrum}
\end{table*}

\subsubsection{$Y_{\rm c}$ states}
There is a rich spectrum of charmonium-like $J^{PC}=1^{--}$ vector states, the so-called $Y$ states.
Below, we discuss our tetraquark model assignments.

Starting from $Y(4008)$, the presence of a broad structure, with mass $4008\pm40^{+114}_{-28}$ MeV and width $226 \pm 44 \pm 87$ MeV, was indicated by Belle in the measured $\pi^+ \pi^- J/\psi$ mass spectrum \cite{Yuan:2007sj}.
However, BaBar did not found the $Y(4008)$ signal in the same $e^+ e^- \rightarrow \pi^+ \pi^- J/\psi$ process \cite{Lees:2012cn}.
Future experiments will give a concluding answer about the $Y(4008)$ existence.
In our tetraquark model calculation, the $Y(4008)$ is interpreted as a $P$-wave scalar diquark-antidiquark bound state.

$Y(4260)$ was discovered by BaBar in $e^+ e^- \rightarrow Y \rightarrow \pi^+ \pi^- J/\psi$ \cite{Choi:2007wga} and then confirmed by CLEO-c \cite{He:2006kg} and Belle \cite{Yuan:2007sj}.
We interpret it as a $P$-wave scalar diquark, axial-vector antidiquark bound state.
In Ref. \cite{Ebert:2008wm} it was described as a $P$-wave scalar diquark-antidiquark bound state, in Ref. \cite{Maiani:2005pe} as the first orbital excitation of a diquark-antidiquark state $cs \bar c \bar s$, but it was also interpreted as a hybrid charmonium in Refs. \cite{Zhu:2005hp,Close:2005iz,Kou:2005gt}. The authors of Refs.~\cite{Wang:2013cya,Wang:2013kra} also interpreted $Y(4260)$ as $\bar{D}D_{1}(2420)$ molecule with a binding energy of $29$ MeV.
Very recently, a possible molecular scenario was discussed in a coupled-channel analysis~\cite{Lu:2017yhl}.

BaBar found evidence of the $Y(4360)$ in $e^+ e^- \rightarrow Y \rightarrow \pi^+ \pi^- \psi(2S)$ \cite{Aubert:2007zz}; later, the $Y(4360)$ was confirmed by Belle, which also found another peak, corresponding to $Y(4660)$ \cite{Wang:2007ea}.
Analogously as in Ref. \cite{Maiani:2014aja}, we interpret $Y(4360)$ and $Y(4660)$ as the second and third radial excitations of $Y(4008)$, respectively.
There are also other possible descriptions for these states, for example $Y(4260)$ and $Y(4360)$ were embedded into the hadro-charmonium picture \cite{Li:2013ssa}, $Y(4260)$, $Y(4360)$ and $Y(4660)$ in a baryonium description \cite{Qiao:2007ce}, while in Ref. \cite{Guo:2008zg} $Y(4660)$ is assumed to be a $f_0(980) \psi(2S)$ bound-state.

Finally, the $Y(4630)$ was seen in $e^+ e^- \rightarrow Y \rightarrow \Lambda_{\rm c} \bar \Lambda_{\rm c}$ by Belle \cite{Pakhlova:2008vn}.
We interpret it as the $Y(4260)$ radial excitation.
In Ref. \cite{Liu:2016sip}, the authors discussed the $Y(4630) \rightarrow \Lambda_{\rm c} \bar \Lambda_{\rm c}$ decay mode in the $^3P_0$ model formalism, under the hypothesis that the $Y(4630)$ is a $1^{--}$ charmonium-like tetraquark.
Because of its peculiar decay mode, $Y(4630)$ was also described as a baryonium state, namely as a $\Lambda_{\rm c} \bar \Lambda_{\rm c}$ bound state \cite{Lee:2011rka}.

\subsection{$s\bar s c \bar c$ tetraquark spectrum}
There are charmonium-like mesons whose decay modes and production mechanisms suggest the presence of $s \bar s$ degrees of freedom in the tetraquark wave function. A typical example is the $X(4140)$, observed in $B \rightarrow K Y(4140)$, with $Y(4140) \rightarrow \phi J/\psi$, by CDF \cite{Aaltonen:2009tz}.
In addition to the $X(4140)$, the CDF Collaboration found evidence of the $X(4274)$ with approximate significance of $3.1 \sigma$ \cite{Aaltonen:2011at}. The related peaks of $J/\psi \phi$ mass structures around $4.3$ GeV were also reported by LHCb, CMS, D0 and BaBar Collaborations~\cite{Aaij:2012pz,Chatrchyan:2013dma,Abazov:2013xda,Lees:2014lra},
which may be the same state as the X(4274).
Very recently, the $X(4140)$ and $X(4274)$ were confirmed by LHCb, which also found evidence of two more structures, the $X(4500)$ and the $X(4700)$ \cite{Aaij:2016iza}.

We interpret the $X(4140)$ as the $s\bar s c \bar c$ counterpart of the $X(3872)$; $X(4500)$ and $X(4700)$ as $0^{++}$ radial excitations of $S$-wave scalar diquark-antidiquark and axial-vector diquark-antidiquark bound states, respectively.
We cannot provide any assignment for the $X(4274)$.

An investigation similar to ours was conducted in Ref. \cite{Lu:2016cwr}. There, the authors studied $s\bar s c \bar c$ tetraquarks within the relativized quark model \cite{Godfrey:1985xj} and discussed possible assignments for $X(4140)$, $X(4274)$, $X(4500)$ and $X(4700)$.
In the $X(4140)$ case, their interpretation coincides with ours.
They also obtain $0^{++}$ radial excitations of $S$-wave scalar diquark-antidiquark and axial-vector diquark-antidiquark bound states characterized by similar energies: one of them can be assigned to $X(4700)$.
They could not accommodate the $X(4274)$.
Stancu calculated the $s\bar s c \bar c$ tetraquark spectrum within a simple quark model with chromomagnetic interaction \cite{Stancu:2009ka}.
She interpreted the $X(4140)$ as the strange partner of the $X(3872)$, but she could not accommodate the other $s\bar s c \bar c$ states, $X(4274)$, $X(4500)$ and $X(4700)$.\footnote{The $X(4500)$ and $X(4700)$ were observed at LHCb in $2016$~\cite{Aaij:2016iza}, and the $X(4274)$ was first observed in $2011$ by CDF with a small significance of $3.1\sigma$~\cite{Aaltonen:2011at}, while Stancu's analysis dates back to 2010.}
In Refs. \cite{Liu:2009ei}, a molecular model description for the $X(4140)$ as $D_{\rm s}^{*+} D_{\rm s}^{*-}$ was proposed.

By using QCD sum rules, the $X(4140)$ and $X(4274)$ were interpreted as $S$-wave $c\bar c s\bar s$ tetraquark states with opposite color structures~\cite{Agaev:2017foq}, and, analogously, the $X(4500)$ and $X(4700)$ as the $D$-wave $c\bar c s\bar s$ tetraquark states with opposite color structures~\cite{Chen:2016oma}.
Maiani {\it et al.} suggested to accommodate $X(4140)$, $X(4274)$, $X(4500)$ and $X(4700)$ within two tetraquark multiplets.
In particular, they suggested that the $X(4500)$ and $X(4700)$ are $2S$ $cs \bar c \bar s$ tetraquark states, the $X(4140)$ the $1^{++}$ ground-state, and that the $X(4274)$ may have $0^{++}$ or $2^{++}$ quantum numbers \cite{Maiani:2016wlq}.

\section{Summary}
We calculated the spectrum of $q\bar q c \bar c$ ($q = u,d$) and $s\bar s c \bar c$ tetraquarks in a relativized diquark model, characterized by one-gluon-exchange (OGE) plus confining potential \cite{Anwar:2017toa}. According to our results, we were able to make some clear assignments, as in the case of $X(3872)$, $Z_{\rm c}(3900)$, $Z_{\rm c}(4020)$, $Y(4008)$, $Z_{\rm c}(4240)$, $Y(4260)$, $Y(4360)$, $Y(4630)$, and $Y(4660)$ in the $q\bar q c \bar c$ sector.
Our intrepretation of the $Z_{\rm c}(4430)$ is dubious, because the experimental mass of the meson falls in the middle of the energy interval between our $2[(1,0)1,0]1$ and $1[(1,0)1,2]1$ tetraquark model predictions of Table \ref{tab:tetraquark-spectrum}, with $J^{PC} = 1^{+-}$.
In the $s\bar s c \bar c$ sector, we could accommodate the $X(4140)$, $X(4500)$ and $X(4700)$.
We could not provide any assignment for the $X(4274)$.
A study of the main decay modes of $XYZ$-type exotics in the diquark model will be important to provide a more precise identification of tetraquark candidates.

Our relativized diquark-antidiquark model results are strongly model dependent.
The possible sources of theoretical uncertainties lie in the choice of the effective Hamiltonian and model parameter fitting procedure, and also in the approximations introduced in the tetraquark wave function.
The latter are strictly related to the possible ways of combinating the quark color representations to obtain a color singlet wave function for the tetraquark.

The next step of our study of fully- and doubly-heavy tetraquarks will be an analysis of the ground-state energies, dominant decay modes and production mechanisms, including estimates of total decay widths and production cross-sections.
More precise experimental data for the exotic meson masses and properties and a detailed comparison between the calculated observables in the main interpretations (tetraquark, molecular model, hadro-quarkonium, and so on) may help to rule out one or more of these pictures.

\begin{acknowledgments}
We would like to thank Profs. Feng-Kun Guo and Bing-Song Zou for several helpful
discussions and suggestions. This work is supported by the National Natural Science
Foundation of China through funds provided to the Sino-German CRC 110 ``Symmetries
and the Emergence of Structure in QCD¡± (Grant No. 11621131001), and by the CAS-TWAS
President's Fellowship for International Ph.D. Students.
\end{acknowledgments}

\begin{appendix}

\section{Classification of tetraquark states}
\label{Classification of tetraquark states}

We report a classification of possible tetraquark states. In the following, we use the notation:
\begin{equation}
	\left| J^{PC} \right\rangle = \left| \left[(S_D,S_{\bar D})_S,L\right]_J \right\rangle  \mbox{ },
\end{equation}
where the diquark, $S_D$, and antidiquark, $S_{\bar D}$, spins are coupled to the total spin, $S$; then, the total spin and the orbital angular momentum, $L$, are coupled to the total angular momentum, $J$.

\begin{itemize}
\item $J^{PC} = 0^{++}$
\begin{subequations}
\begin{equation}
	\left| 0^{++} \right\rangle = \left| \left[(0,0)_0,0\right]_0 \right\rangle  \mbox{ } (^1S_0)  \mbox{ },
\end{equation}
\begin{equation}
	\left| 0^{++} \right\rangle = \left| \left[(1,1)_0,0\right]_0 \right\rangle  \mbox{ } (^1S_0)
	\mbox{ },
\end{equation}
\begin{equation}
	\left| 0^{++} \right\rangle = \left| \left[(1,1)_2,2\right]_0 \right\rangle  \mbox{ } (^5D_0)
	\mbox{ }.
\end{equation}
\end{subequations}

\item $J^{PC} = 1^{++}$
\begin{subequations}
\begin{equation}
	\begin{array}{rcl}
	\left| 1^{++} \right\rangle & = & \frac{1}{\sqrt 2} \left[ \left| \left[ (0,1)_1,0\right]_1 \right\rangle  \right. \\
	& + & \left. \left| \left[(1,0)_1,0\right]_1 \right\rangle \right]	\mbox{ } (^3S_1)  \mbox{ },
	\end{array}
\end{equation}
\begin{equation}
	\begin{array}{rcl}
	\left| 1^{++} \right\rangle & = & \frac{1}{\sqrt 2} \left[ \left| \left[(0,1)_1,2\right]_1 \right\rangle  \right. \\
	& + & \left. \left| \left[ (1,0)_1,2\right]_1 \right\rangle \right]	\mbox{ } (^3D_1)  \mbox{ },
	\end{array}
\end{equation}
\begin{equation}
	\left| 1^{++} \right\rangle = \left| \left[(1,1)_2,2\right]_1 \right\rangle  \mbox{ } (^5D_1)
	\mbox{ }.
\end{equation}
\end{subequations}

\item $J^{PC} = 1^{+-}$
\begin{subequations}
\begin{equation}
	\left| 1^{+-} \right\rangle = \left| \left[(1,1)_1,0\right]_1 \right\rangle  \mbox{ } (^3S_1)
	\mbox{ },
\end{equation}
\begin{equation}
	\begin{array}{rcl}
	\left| 1^{+-} \right\rangle & = & \frac{1}{\sqrt 2} \left[ \left| \left[(0,1)_1,0\right]_1 \right\rangle  \right. \\
	& - & \left. \left| \left[(1,0)_1,0\right]_1 \right\rangle \right]	\mbox{ } (^3S_1)  \mbox{ },
	\end{array}
\end{equation}
\begin{equation}
	\begin{array}{rcl}
	\left| 1^{+-} \right\rangle & = & \frac{1}{\sqrt 2} \left[\left| \left[(0,1)_1,2\right]_1 \right\rangle  \right. \\
	& - & \left. \left| \left[(1,0)_1,2\right]_1 \right\rangle \right]	\mbox{ } (^3D_1)  \mbox{ },
	\end{array}
\end{equation}
\begin{equation}
	\left| 1^{+-} \right\rangle = \left| \left[(1,1)_1,2\right]_1 \right\rangle  \mbox{ } (^3D_1)
	\mbox{ }.
\end{equation}
\end{subequations}

\item $J^{PC} = 0^{-+}$
\begin{subequations}
\begin{equation}
	\begin{array}{rcl}
	\left| 0^{-+} \right\rangle & = & \frac{1}{\sqrt 2} \left[ \left| \left[(0,1)_1,1\right]_0 \right\rangle  \right. \\
	& + & \left. \left| \left[(1,0)_1,1\right]_0 \right\rangle \right]	\mbox{ } (^3P_0)  \mbox{ },
	\end{array}
\end{equation}
\begin{equation}
	\left| 0^{-+} \right\rangle = \left| \left[(1,1)_1,1\right]_0 \right\rangle  \mbox{ } (^3P_0)
	\mbox{ }.
\end{equation}
\end{subequations}

\item $J^{PC} = 0^{--}$
\begin{equation}
	\begin{array}{rcl}
	\left| 0^{--} \right\rangle & = & \frac{1}{\sqrt 2} \left[ \left| \left[(0,1)_1,1\right]_0 \right\rangle  \right. \\
	& - & \left. \left| \left[(1,0)_1,1\right]_0 \right\rangle \right]	\mbox{ } (^3P_0)  \mbox{ }.
	\end{array}
\end{equation}

\item $J^{PC} = 1^{--}$
\begin{subequations}
\begin{equation}
	\left| 1^{--} \right\rangle = \left| \left[(0,0)_0,1\right]_1 \right\rangle  \mbox{ } (^1P_1)
	\mbox{ },
\end{equation}
\begin{equation}
	\left| 1^{--} \right\rangle = \left| \left[(1,1)_0,1\right]_1 \right\rangle  \mbox{ } (^1P_1)
	\mbox{ },
\end{equation}
\begin{equation}
	\begin{array}{rcl}
	\left| 1^{--} \right\rangle & = & \frac{1}{\sqrt 2} \left[ \left| \left[(0,1)_1,1\right]_1 \right\rangle  \right. \\
	& - & \left. \left| \left[(1,0)_1,1\right]_1 \right\rangle \right]	\mbox{ } (^3P_1)  \mbox{ },
	\end{array}
\end{equation}
\begin{equation}
	\left| 1^{--} \right\rangle = \left| \left[(1,1)_2,1\right]_1 \right\rangle  \mbox{ } (^5P_1)
	\mbox{ }.
\end{equation}
\end{subequations}

\end{itemize}
\end{appendix}


\begin{thebibliography}{}

\bibitem{Greiner:1989eu}
  W.~Greiner and B.~Muller,
  ``Theoretical physics. Vol. 2: Quantum mechanics. Symmetries'',
  Springer (1989).

\bibitem{Richard:1992uk}
  J.~M.~Richard,
  Phys.\ Rept.\  {\bf 212}, 1 (1992).

\bibitem{Buchmuller:1992zf}
  W.~Buchmuller (Ed.),
  ``Quarkonia'',
  North-Holland (1992).

\bibitem{Capstick:2000qj}
  S.~Capstick and W.~Roberts,
  Prog.\ Part.\ Nucl.\ Phys.\  {\bf 45}, S241 (2000).

\bibitem{Ablikim}
  M. Ablikim {\it et al.} [BESIII Collaboration],
  Phys.\ Rev.\ Lett. {\bf 110}, 252001 (2013).

\bibitem{Liu}
  Z. Q. Liu {\it et al.} [Belle Collaboration],
  Phys.\ Rev.\ Lett. {\bf 110}, 252002 (2013).

\bibitem{Ablikim:2013wzq}
  M.~Ablikim {\it et al.} [BESIII Collaboration],
  Phys.\ Rev.\ Lett.\  {\bf 111}, 242001 (2013).

\bibitem{Ablikim:2013emm}
  M.~Ablikim {\it et al.} [BESIII Collaboration],
  Phys.\ Rev.\ Lett.\  {\bf 112}, 132001 (2014).
	
\bibitem{Bondar}
  A. Bondar {\it et al.} [Belle Collaboration],
  Phys.\ Rev.\ Lett. {\bf 108}, 122001 (2012).		
	
\bibitem{Choi:2003ue}
  S.~K.~Choi {\it et al.} [Belle Collaboration],
  Phys.\ Rev.\ Lett.\  {\bf 91}, 262001 (2003).	

\bibitem{Jaffe:1976ih}
  R.~L.~Jaffe,
  Phys.\ Rev.\ D {\bf 15}, 281 (1977).

\bibitem{Barbour:1979qi}
  I.~M.~Barbour and D.~K.~Ponting,
  Z.\ Phys.\ C {\bf 5}, 221 (1980);
  I.~M.~Barbour and J.~P.~Gilchrist,
  Z.\ Phys.\ C {\bf 7}, 225 (1981)
  Erratum: [Z.\ Phys.\ C {\bf 8}, 282 (1981)].

\bibitem{Weinstein:1983gd}
  J.~D.~Weinstein and N.~Isgur,
  Phys.\ Rev.\ D {\bf 27}, 588 (1983).	

\bibitem{SilvestreBrac:1993ss}
  B.~Silvestre-Brac and C.~Semay,
  Z.\ Phys.\ C {\bf 57}, 273 (1993).

\bibitem{Brink:1998as}
  D.~M.~Brink and F.~Stancu,
  Phys.\ Rev.\ D {\bf 57}, 6778 (1998).

\bibitem{Maiani:2004vq}
  L.~Maiani, F.~Piccinini, A.~D.~Polosa and V.~Riquer,
  Phys.\ Rev.\ Lett.\  {\bf 93}, 212002 (2004);
  Phys.\ Rev.\ D {\bf 71}, 014028 (2005);
  {\bf 72}, 031502 (2005);
  L.~Maiani, V.~Riquer, R.~Faccini, F.~Piccinini, A.~Pilloni and A.~D.~Polosa,
  Phys.\ Rev.\ D {\bf 87}, no. 11, 111102 (2013).

\bibitem{Barnea:2006sd}
  N.~Barnea, J.~Vijande and A.~Valcarce,
  Phys.\ Rev.\ D {\bf 73}, 054004 (2006).

\bibitem{Santopinto:2006my}
  E.~Santopinto and G.~Galat\`a,
  Phys.\ Rev.\ C {\bf 75}, 045206 (2007).

\bibitem{Ebert:2008wm}
  D.~Ebert, R.~N.~Faustov, V.~O.~Galkin and W.~Lucha,
  Phys.\ Rev.\ D {\bf 76}, 114015 (2007);
  D.~Ebert, R.~N.~Faustov and V.~O.~Galkin,
  Phys.\ Atom.\ Nucl.\  {\bf 72}, 184 (2009).

\bibitem{Deng:2014gqa}
  C.~Deng, J.~Ping and F.~Wang,
  Phys.\ Rev.\ D {\bf 90}, 054009 (2014)

\bibitem{Zhao:2014qva}
  L.~Zhao, W.~Z.~Deng and S.~L.~Zhu,
  Phys.\ Rev.\ D {\bf 90}, 094031 (2014).

\bibitem{Anwar:2017toa}
  M.~N.~Anwar, J.~Ferretti, F.~K.~Guo, E.~Santopinto and B.~S.~Zou,
  arXiv:1710.02540.

\bibitem{Dubynskiy:2008mq}
  S.~Dubynskiy and M.~B.~Voloshin,
  Phys.\ Lett.\ B {\bf 666}, 344 (2008).

\bibitem{Voloshin:2013dpa}
  M.~B.~Voloshin,
  Phys.\ Rev.\ D {\bf 87}, 091501 (2013).

\bibitem{Li:2013ssa}
  X.~Li and M.~B.~Voloshin,
  Mod.\ Phys.\ Lett.\ A {\bf 29}, 1450060 (2014).

\bibitem{Wang:2013kra}
  Q.~Wang, M.~Cleven, F.~K.~Guo, C.~Hanhart, U.~G.~Mei\ss ner, X.~G.~Wu and Q.~Zhao,
  Phys.\ Rev.\ D {\bf 89}, 034001 (2014);
  M.~Cleven, F.~K.~Guo, C.~Hanhart, Q.~Wang and Q.~Zhao,
  Phys.\ Rev.\ D {\bf 92}, 014005 (2015).

\bibitem{Brambilla:2015rqa}
  N.~Brambilla, G.~Krein, J.~Tarr¨²s Castell¨¤ and A.~Vairo,
  Phys.\ Rev.\ D {\bf 93}, 054002 (2016).

\bibitem{Panteleeva:2018ijz}
  J.~Y.~Panteleeva, I.~A.~Perevalova, M.~V.~Polyakov and P.~Schweitzer,
  arXiv:1802.09029.

\bibitem{Ferretti:2018kzy}
  J.~Ferretti,
  Phys.\ Lett.\ B {\bf 782}, 702 (2018).

\bibitem{Weinstein:1990gu}
  J.~D.~Weinstein and N.~Isgur,
  Phys.\ Rev.\ D {\bf 41}, 2236 (1990).

\bibitem{Manohar:1992nd}
  A.~V.~Manohar and M.~B.~Wise,
  Nucl.\ Phys.\ B {\bf 399}, 17 (1993).

\bibitem{Tornqvist:1993ng}
  N.~A.~T\"ornqvist,
  Z.\ Phys.\ C {\bf 61}, 525 (1994);
  Phys.\ Lett.\ B {\bf 590}, 209 (2004).

\bibitem{Swanson:2003tb}
  E.~S.~Swanson,
  Phys.\ Lett.\ B {\bf 588}, 189 (2004);
  {\bf 598}, 197 (2004).

\bibitem{Hanhart:2007yq}
  C.~Hanhart, Y.~S.~Kalashnikova, A.~E.~Kudryavtsev and A.~V.~Nefediev,
  Phys.\ Rev.\ D {\bf 76}, 034007 (2007).

\bibitem{Thomas:2008ja}
  C.~E.~Thomas and F.~E.~Close,
  Phys.\ Rev.\ D {\bf 78}, 034007 (2008).

\bibitem{Baru:2011rs}
   V.~Baru, A.~A.~Filin, C.~Hanhart, Y.~S.~Kalashnikova, A.~E.~Kudryavtsev and A.~V.~Nefediev,
  Phys.\ Rev.\ D {\bf 84}, 074029 (2011).

\bibitem{Valderrama:2012jv}
  M.~P.~Valderrama,
  Phys.\ Rev.\ D {\bf 85}, 114037 (2012).

\bibitem{Guo:2013sya}
  F.~K.~Guo, C.~Hidalgo-Duque, J.~Nieves and M.~P.~Valderrama,
  Phys.\ Rev.\ D {\bf 88}, 054007 (2013).

\bibitem{Kang:2016jxw}
  X.~W.~Kang and J.~A.~Oller,
  Eur.\ Phys.\ J.\ C {\bf 77}, 399 (2017).

\bibitem{Heikkila:1983wd}
  K.~Heikkila, S.~Ono and N.~A.~Tornqvist,
  Phys.\ Rev.\ D {\bf 29}, 110 (1984)
  Erratum: [Phys.\ Rev.\ D {\bf 29}, 2136 (1984)].
 		
\bibitem{Pennington:2007xr}
  M.~R.~Pennington and D.~J.~Wilson,
  Phys.\ Rev.\ D {\bf 76}, 077502 (2007).

\bibitem{Danilkin:2010cc}
  I.~V.~Danilkin and Y.~A.~Simonov,
  Phys.\ Rev.\ Lett.\  {\bf 105}, 102002 (2010).

\bibitem{bottomonium}
  J.~Ferretti, G.~Galat\`a and E.~Santopinto,
  Phys.\ Rev.\ C {\bf 88}, 015207 (2013);
  Phys.\ Rev.\ D {\bf 90}, 054010 (2014);
  J.~Ferretti and E.~Santopinto,
  Phys.\ Rev.\ D {\bf 90}, 094022 (2014);
  arXiv:1806.02489.

\bibitem{Lu:2016mbb}
  Y.~Lu, M.~N.~Anwar and B.~S.~Zou,
  Phys.\ Rev.\ D {\bf 94}, 034021 (2016);
  Phys.\ Rev.\ D {\bf 95}, 034018 (2017);
  Phys.\ Rev.\ D {\bf 96}, 114022 (2017).

\bibitem{Guo:2014iya}
  F.~K.~Guo, C.~Hanhart, Q.~Wang and Q.~Zhao,
  Phys.\ Rev.\ D {\bf 91}, 051504 (2015).

\bibitem{Szczepaniak:2015eza}
  A.~P.~Szczepaniak,
  Phys.\ Lett.\ B {\bf 747}, 410 (2015).

\bibitem{Liu:2015taa}
  X.~H.~Liu, M.~Oka and Q.~Zhao,
  Phys.\ Lett.\ B {\bf 753}, 297 (2016).	

\bibitem{Esposito:2016noz}
  A.~Esposito, A.~Pilloni and A.~D.~Polosa,
  Phys.\ Rept.\  {\bf 668}, 1 (2016).

\bibitem{Ali:2017jda}
  A.~Ali, J.~S.~Lange and S.~Stone,
  Prog.\ Part.\ Nucl.\ Phys.\  {\bf 97}, 123 (2017).

\bibitem{Olsen:2017bmm}
  S.~L.~Olsen, T.~Skwarnicki and D.~Zieminska,
  Rev.\ Mod.\ Phys.\  {\bf 90}, 015003 (2018).

\bibitem{Karliner:2017qhf}
  M.~Karliner, J.~L.~Rosner and T.~Skwarnicki,
  arXiv:1711.10626.

\bibitem{Chen:2016qju}
  H.~X.~Chen, W.~Chen, X.~Liu and S.~L.~Zhu,
  Phys.\ Rept.\  {\bf 639}, 1 (2016).

\bibitem{Guo:2017jvc}
  F.~K.~Guo, C.~Hanhart, U.~G.~Meißner, Q.~Wang, Q.~Zhao and B.~S.~Zou,
  Rev.\ Mod.\ Phys.\  {\bf 90}, 015004 (2018).

\bibitem{Lebed:2016hpi}
  R.~F.~Lebed, R.~E.~Mitchell and E.~S.~Swanson,
  Prog.\ Part.\ Nucl.\ Phys.\  {\bf 93}, 143 (2017).

\bibitem{Berezhnoy:2011xn}
  A.~V. Berezhnoy, A.~V. Luchinsky, and A.~A. Novoselov,
  Phys. Rev. {\bf D86}, 034004 (2012).

\bibitem{Chen:2016jxd}
  W.~Chen, H.-X. Chen, X.~Liu, T.~G. Steele, and S.-L. Zhu,
  Phys.\ Lett.\ B {\bf 773}, 247 (2017).

\bibitem{Karliner:2016zzc}
  M.~Karliner, S.~Nussinov, and J.~L. Rosner,
  Phys. Rev. {\bf D95}, 034011 (2017).

\bibitem{Richard:2017vry}
  J.-M. Richard, A.~Valcarce, and J.~Vijande,
  Phys. Rev. {\bf D95}, 054019 (2017).

\bibitem{Wang:2017jtz}
  Z.-G. Wang,
  Eur.\ Phys.\ J.\ C {\bf 77}, 432 (2017).

\bibitem{Lipkin:1986dw}
  H.~J.~Lipkin,
  Phys.\ Lett.\ B {\bf 172}, 242 (1986).

\bibitem{Ader:1981db}
  J.~P.~Ader, J.~M.~Richard and P.~Taxil,
  Phys.\ Rev.\ D {\bf 25}, 2370 (1982).

\bibitem{Karliner:2017qjm}
  M.~Karliner and J.~L.~Rosner,
  Phys.\ Rev.\ Lett.\  {\bf 119}, 202001 (2017).

\bibitem{Eichten:2017ffp}
  E.~J.~Eichten and C.~Quigg,
  Phys.\ Rev.\ Lett.\  {\bf 119}, 202002 (2017).

\bibitem{Czarnecki:2017vco}
  A.~Czarnecki, B.~Leng and M.~B.~Voloshin,
  Phys.\ Lett.\ B {\bf 778}, 233 (2018).

\bibitem{Luo:2017eub}
  S.~Q.~Luo, K.~Chen, X.~Liu, Y.~R.~Liu and S.~L.~Zhu,
  Eur.\ Phys.\ J.\ C {\bf 77}, 709 (2017).

\bibitem{Godfrey:1985xj}
  S.~Godfrey and N.~Isgur,
  Phys.\ Rev.\ D {\bf 32}, 189 (1985).

\bibitem{Capstick:1986bm}
  S.~Capstick and N.~Isgur,
  Phys.\ Rev.\ D {\bf 34}, 2809 (1986).

\bibitem{Capstick:1992th}
  S.~Capstick and W.~Roberts,
  Phys.\ Rev.\ D {\bf 47}, 1994 (1993).

\bibitem{Bijker:1994yr}
  R.~Bijker, F.~Iachello and A.~Leviatan,
  Annals Phys.\  {\bf 236}, 69 (1994).

\bibitem{Ferraris:1995ui}
  M.~Ferraris, M.~M.~Giannini, M.~Pizzo, E.~Santopinto and L.~Tiator,
  Phys.\ Lett.\ B {\bf 364}, 231 (1995).

\bibitem{Ferretti:2011zz}
  J.~Ferretti, A.~Vassallo and E.~Santopinto,
  Phys.\ Rev.\ C {\bf 83}, 065204 (2011);
  E.~Santopinto and J.~Ferretti,
  Phys.\ Rev.\ C {\bf 92}, 025202 (2015);
  M.~De Sanctis, J.~Ferretti, E.~Santopinto and A.~Vassallo,
  Eur.\ Phys.\ J.\ A {\bf 52}, no. 5, 121 (2016).

\bibitem{Ferretti:2015ada}
  J.~Ferretti, R.~Bijker, G.~Galat\`a, H.~Garc\'ia-Tecocoatzi and E.~Santopinto,
  Phys.\ Rev.\ D {\bf 94}, 074040 (2016).

\bibitem{Jaffe:2004ph}
  R.~L.~Jaffe,
  Phys.\ Rept.\  {\bf 409}, 1 (2005).

\bibitem{Lichtenberg:1996fi}
  D.~B.~Lichtenberg, R.~Roncaglia and E.~Predazzi,
  In *Turin 1996, Diquarks III* 146-155
  [hep-ph/9611428].		

\bibitem{Celmaster:1977vh}
  W.~Celmaster, H.~Georgi and M.~Machacek,
  Phys.\ Rev.\ D {\bf 17}, 879 (1978).

\bibitem{Gradshteyn-Ryzhik}
  I. S. Gradshteyn and I. M. Ryzhik,
  ``Table of Integrals, Series, and Products'',
  Academic Press.

\bibitem{Nakamura:2010zzi}
  C. Patrignani {\it et al.} [Particle Data Group],
  Chin. Phys. C {\bf 40}, 100001 (2016).

\bibitem{Aubert:2004ns}
  B.~Aubert {\it et al.} [BaBar Collaboration],
  Phys.\ Rev.\ D {\bf 71}, 071103 (2005).

\bibitem{Acosta:2003zx}
  D.~Acosta {\it et al.} [CDF Collaboration],
  Phys.\ Rev.\ Lett.\  {\bf 93}, 072001 (2004).

\bibitem{Abazov:2004kp}
  V.~M.~Abazov {\it et al.} [D0 Collaboration],
  Phys.\ Rev.\ Lett.\  {\bf 93}, 162002 (2004).

\bibitem{Aaij:2014jqa}
  R.~Aaij {\it et al.} [LHCb Collaboration],
  Phys.\ Rev.\ Lett.\  {\bf 112}, 222002 (2014).

\bibitem{Aaij:2015zxa}
  R.~Aaij {\it et al.} [LHCb Collaboration],
  Phys.\ Rev.\ D {\bf 92}, 112009 (2015).

\bibitem{Choi:2007wga}
  S.~K.~Choi {\it et al.} [Belle Collaboration],
  Phys.\ Rev.\ Lett.\  {\bf 100}, 142001 (2008).
  
\bibitem{Agaev:2017tzv}
  S.~S.~Agaev, K.~Azizi and H.~Sundu,
  Phys.\ Rev.\ D {\bf 96}, 034026 (2017);
  R.~Zhu,
  Phys.\ Rev.\ D {\bf 94}, 054009 (2016).
  
\bibitem{Yuan:2007sj}
  C.~Z.~Yuan {\it et al.} [Belle Collaboration],
  Phys.\ Rev.\ Lett.\  {\bf 99}, 182004 (2007).

\bibitem{Lees:2012cn}
  J.~P.~Lees {\it et al.} [BaBar Collaboration],
  Phys.\ Rev.\ D {\bf 86}, 051102 (2012).

\bibitem{He:2006kg}
  Q.~He {\it et al.} [CLEO Collaboration],
  Phys.\ Rev.\ D {\bf 74}, 091104 (2006).

\bibitem{Maiani:2005pe}
  L.~Maiani, V.~Riquer, F.~Piccinini and A.~D.~Polosa,
  Phys.\ Rev.\ D {\bf 72}, 031502 (2005).

\bibitem{Zhu:2005hp}
  S.~L.~Zhu,
  Phys.\ Lett.\ B {\bf 625}, 212 (2005).

\bibitem{Close:2005iz}
  F.~E.~Close and P.~R.~Page,
  Phys.\ Lett.\ B {\bf 628}, 215 (2005).

\bibitem{Kou:2005gt}
  E.~Kou and O.~Pene,
  Phys.\ Lett.\ B {\bf 631}, 164 (2005).

\bibitem{Wang:2013cya}
  Q.~Wang, C.~Hanhart and Q.~Zhao,
  Phys.\ Rev.\ Lett.\  {\bf 111}, 132003 (2013).

\bibitem{Lu:2017yhl}
  Y.~Lu, M.~N.~Anwar and B.~S.~Zou,
  Phys.\ Rev.\ D {\bf 96}, 114022 (2017).

\bibitem{Aubert:2007zz}
  B.~Aubert {\it et al.} [BaBar Collaboration],
  Phys.\ Rev.\ Lett.\  {\bf 98}, 212001 (2007).

\bibitem{Wang:2007ea}
  X.~L.~Wang {\it et al.} [Belle Collaboration],
  Phys.\ Rev.\ Lett.\  {\bf 99}, 142002 (2007).

\bibitem{Maiani:2014aja}
  L.~Maiani, F.~Piccinini, A.~D.~Polosa and V.~Riquer,
  Phys.\ Rev.\ D {\bf 89}, 114010 (2014).

\bibitem{Qiao:2007ce}
  C.~F.~Qiao,
  J.\ Phys.\ G {\bf 35}, 075008 (2008).

\bibitem{Guo:2008zg}
  F.~K.~Guo, C.~Hanhart and U.~G.~Meissner,
  Phys.\ Lett.\ B {\bf 665}, 26 (2008).

\bibitem{Pakhlova:2008vn}
  G.~Pakhlova {\it et al.} [Belle Collaboration],
  Phys.\ Rev.\ Lett.\  {\bf 101}, 172001 (2008).

\bibitem{Liu:2016sip}
  X.~Liu, H.~W.~Ke, X.~Liu and X.~Q.~Li,
  Eur.\ Phys.\ J.\ C {\bf 76}, 549 (2016).

\bibitem{Lee:2011rka}
  N.~Lee, Z.~G.~Luo, X.~L.~Chen and S.~L.~Zhu,
  Phys.\ Rev.\ D {\bf 84}, 014031 (2011).

\bibitem{Aaltonen:2009tz}
  T.~Aaltonen {\it et al.} [CDF Collaboration],
  Phys.\ Rev.\ Lett.\  {\bf 102}, 242002 (2009).

\bibitem{Aaltonen:2011at}
  T.~Aaltonen {\it et al.} [CDF Collaboration],
  Mod.\ Phys.\ Lett.\ A {\bf 32}, 1750139 (2017).

\bibitem{Aaij:2012pz}
  R.~Aaij {\it et al.} [LHCb Collaboration],
  Phys.\ Rev.\ D {\bf 85}, 091103 (2012).

\bibitem{Chatrchyan:2013dma}
  S.~Chatrchyan {\it et al.} [CMS Collaboration],
  Phys.\ Lett.\ B {\bf 734}, 261 (2014).

\bibitem{Abazov:2013xda}
  V.~M.~Abazov {\it et al.} [D0 Collaboration],
  Phys.\ Rev.\ D {\bf 89}, 012004 (2014).

\bibitem{Lees:2014lra}
  J.~P.~Lees {\it et al.} [BaBar Collaboration],
  Phys.\ Rev.\ D {\bf 91}, 012003 (2015).

\bibitem{Aaij:2016iza}
  R.~Aaij {\it et al.} [LHCb Collaboration],
  Phys.\ Rev.\ Lett.\  {\bf 118}, 022003 (2017);
  Phys.\ Rev.\ D {\bf 95}, 012002 (2017).

\bibitem{Lu:2016cwr}
  Q.~F.~L\"u and Y.~B.~Dong,
  Phys.\ Rev.\ D {\bf 94}, 074007 (2016).

\bibitem{Stancu:2009ka}
  F.~Stancu,
  J.\ Phys.\ G {\bf 37}, 075017 (2010).

\bibitem{Liu:2009ei}
  X.~Liu and S.~L.~Zhu,
  Phys.\ Rev.\ D {\bf 80}, 017502 (2009)
  Erratum: [Phys.\ Rev.\ D {\bf 85}, 019902 (2012)];
  X.~Liu, Z.~G.~Luo, Y.~R.~Liu and S.~L.~Zhu,
  Eur.\ Phys.\ J.\ C {\bf 61}, 411 (2009).

\bibitem{Chen:2016oma}
  H.~X.~Chen, E.~L.~Cui, W.~Chen, X.~Liu and S.~L.~Zhu,
  Eur.\ Phys.\ J.\ C {\bf 77}, 160 (2017).

\bibitem{Agaev:2017foq}
  S.~S.~Agaev, K.~Azizi and H.~Sundu,
  Phys.\ Rev.\ D {\bf 95}, 114003 (2017).

\bibitem{Maiani:2016wlq}
  L.~Maiani, A.~D.~Polosa and V.~Riquer,
  Phys.\ Rev.\ D {\bf 94}, 054026 (2016).
			
\end{thebibliography}
\end{document}